# High transport currents in mechanically reinforced MgB$_2$ wires

W Goldacker, SI Schlachter, S Zimmer, H Reiner

Forschungszentrum Karlsruhe - Institut für Technische Physik (ITP), P.O. Box 3640, D-76021 Karlsruhe, Germany.

E-mail: wilfried.goldacker@itp.fzk.de

**Abstract**
We prepared and characterized monofilamentary MgB$_2$ wires with a mechanically reinforced composite sheath of Ta(Nb)/Cu/steel, which leads to dense filaments and correspondingly high transport currents up to $J_c$ = 10$^5$ Acm$^{-2}$ at 4.2 K, self field. The reproducibility of the measured transport currents was excellent and not depending on the wire diameter. Using different precursors, commercial reacted powder or an unreacted Mg/B powder mixture, a strong influence on the pinning behaviour and the irreversibility field was observed. The critical transport current density showed a nearly linear temperature dependency for all wires being still 52 kAcm$^{-2}$ at 20 K and 23 kAcm$^{-2}$ at 30 K. Detailed data for $J_c(B,T)$ and $T_c(B)$ were measured.

**Introduction**

The recent discovery of superconductivity in MgB$_2$ with a $T_c$ of 39 K [1,2] gives rise to hope of a new superconductor system suitable for applications, besides the well established NbTi and Nb$_3$Sn conductor materials. The reported upper critical field $\mu_0 H_{c2}(0)$ = 17 – 18 T is well below the values obtained for Nb$_3$Sn wires and the irreversibility field was found to be approximately 0.5·$\mu_0 H_{c2}(0)$ [3-6]. New aspects raised regarding recent thin film investigations on doped materials which showed a very promising anisotropy of the upper critical field up to 39 T [7]. A possible improvement of wires or tapes seems to be achievable developing conductors with textured MgB$_2$. Different efforts were made so far to prepare wires or tapes. Applying a heat treatment to the sample requires a protective sheath like Nb [8], or Ta [9] to avoid reaction with the decomposition products, e.g. elementary Mg. Several other sheath compositions like Cu/Fe [10] were applied with similar results. Also the precursor was varied from commercial pre-reacted powders to non-reacted powder mixtures of Mg/B [11]. A tape conductor without any further heat treatment was processed with a Ag, Cu or Ni sheath, which gives a good densification due to the high rolling pressure [12]. Critical transport current densities reported so far reached values of up to 70 kAcm$^{-2}$ [8] in heat treated conductors and even 100 kAcm$^{-2}$ in the as rolled tapes [12] at 4.2 K and self field. Much higher current densities exceeding 1 MAcm$^{-2}$, measured magnetically on films [13,14], demonstrate the potential for further improving the current carrying behaviour of wires and tapes.

For the design of MgB$_2$ wires several aspects need to be considered. First, the geometrical shape of a wire is in principal more appropriate compared to a tape with respect to the standard coil winding process. If the anisotropy of $\mu_0 H_{c2}$ which was found in textured MgB$_2$ thin films to reach a factor of two [7], gives the chance to improve critical transport currents, especially in oriented background fields, a tape geometry might become more important or even necessary, in combination with a more sophisticated in situ phase processing. - One further question is the composition and constitution of the precursor. Mainly two possibilities presently exist. The use of commercial MgB$_2$ powders requires a powder in tube wire (PIT) process similar to the BSCCO system, with the disadvantage of a non ductile deformation of the filament. Second, the use of a mixture of Mg and B elementary powders has the advantage of more or less ductile deformation properties but the disadvantage of less fine powder particles and a possibly less homogeneous $MgB_2$ formation. Both methods were applied in our investigation.

From work on pressed and heat treated bulk samples, the very poor sinter property of the material was recognized, and therefore a heat treatment of the wires at temperatures above that of the $MgB_2$ decomposition, then falling through the recovery range of the $MgB_2$ phase upon slow cooling, is considered to be a necessary requirement to get a dense phase.

Regarding the boundary conditions, the next aspect, the choice of the sheath being in contact with the superconductor, becomes of crucial importance. During decomposition of $MgB_2$ upon heating, elementary Mg forms parallel to the formation of higher borides, which reacts quantitatively with the oxygen being present. The reactivity of elementary Mg with many metals restricts the choice of the 'first wall material'. From phase diagram considerations, a selection of suitable sheath materials was discussed. We chose Nb and Ta, since they are sufficiently ductile for high deformation ratios, are not reactive with Mg and can bond elementary oxygen. A possible formation of a reaction layer forming borides or oxides was taken into account.

Since the wire composite undergoes a heat treatment between RT and about 900°C, the match of the thermal expansion coefficient of the sheath components and the superconducting filament has to be considered. For a favourable combination of materials, where the sheath has a significant larger thermal expansion compared to the superconductor, a compressive pre-strain and pre-stress builds up during the cooling ramp from the reaction temperature below the temperature where the wire components become elastic. The heating ramp plays a negligible role when reaching a temperature level where the working strains are released and the components are in a plastic state. This quite general behaviour was established by means of high temperature neutron diffraction experiments on NbSn wires with different composite structures measuring the lattice strains with temperature[15]. In standard NbSn wires, lattice strains occurred below 150 - 350°C, in steel reinforced wires, the onset was significantly shifted to higher temperatures of about 500°C. Regarding linear thermal expansion coefficients at RT for materials used in $MgB_2$ wires, given in Tab.1, illustrates the problem to design a mechanical stable sheath in this case. An approximate thermal expansion coefficient for $MgB_2$ polycrystalline untextured bulk (see table 1) was calculated from the temperature dependence of the unit cell volume at 300 K , measured by means of neutron diffraction in ref.[16]. This value is between that of Nb and the values of Fe, Ni, which are 40-50% higher. Copper and steel in this selection have much higher values being twice that of $MgB_2$. The use of Ta, Nb sheaths as single sheath material is therefore very unfavourable, a compressive pre-strain of the $MgB_2$ filament cannot be achieved and filament cracks after cooling may occur. Having selected Ta or Nb as first wall material, the combination with an outer Cu/Steel sheath layer is therefore necessary for mechanical reinforcement and to realise a pre-compression of the filament. For this aspect we took profit of earlier experience with Chevrel wires, where a very similar situation with a comparable expansion coefficient of $9.4 \times 10^{-6}$ $K^{-1}$ was given and an equivalent composite concept worked very well and successful [17,18]. We expect that also in the $MgB_2$ wires, the mechanical stabilisation becomes effective during the cooling ramp at high temperatures (500-600°C) close to the reformation temperature of the phase and material densification due to compressive stress already occurs.

# Experimental

## Preparation of the wires

In the first concept, commercial $MgB_2$ powders (Alfa Aesar) with about 2% MgO (from X ray spectrum) secondary phase - obviously varying with production batch - were used. In the second alternative route, powder mixtures of Mg (< 300 mesh, 99 % purity) and B (< 200 mesh, 99 % purity) with 5 wt% over-stoichiometry of Mg were used. The over-stoichiometry of Mg was chosen to compensate the non avoidable formation of MgO, which occurs from reaction with surface bound oxygen of the Mg ingot. The powders were mixed for several hours and were not especially ground before filling the sheath tubes. The wires made from elementary powders are named "in-situ" wires with respect to the $MgB_2$ formation during the heat treatment of the final wire.

The sheaths being applied were Ta and Nb tubes (10mm diam., 1mm wall thickness) inside a Cu tube of 12 mm outer diameter. The precursor powder was filled in and pressed moderately. Deformation was made by swaging and drawing to 1.6 mm wire diameter. At this stage the wire was introduced in a stainless steel tube of 2.5 mm diam., 0.35 mm wall thickness, and then deformed to the final wire diameters of 0.86, 1.18 and 1.58 mm diameter. The necessary amount of stainless steel was estimated roughly from the analysis made on Chevrel phase wires in refs. [17,18]. A similar systematic work, varying the steel content was not done so far, but is a very important goal of the ongoing work. Wire lengths of 5 - 8 m were produced. All wires had quite regular cross sections shown as an example in Fig.1. The final heat treatment was performed heating up to 900°C, which is well above the phase decomposition, followed by subsequent cooling with 20 °C/hr to 650°C and then furnace cooling to RT. The annealing atmosphere was $Ar/H_2(5\%)$ to avoid oxidation of the steel sheath. So far no systematic optimization of the annealing treatment was performed.

## Transport critical current and $T_c$ measurements

For critical current measurements the wire ends of typically 40 mm long samples were prepared by grinding, electrolytic coating with Cu and soldering with low temperature solder to the sample holder. For nearly all $I_c$ measurements we observed a small contact resistance. For some not accurately prepared samples we observed current sharing effects, indicating that a resistive layer between superconductor and sheath has formed due to some chemical reaction.

The critical current measurements were performed in a flow cryostat or LHe bath with the option to apply a background field up to 0.7 T or 14 T, depending on the used experimental equipment. Also long samples of 0.4 and 1 m length were prepared by the wind & react technique in a coil shape of 33 or 40 mm diameter (see Fig.2). The $I_c$ criterion used was generally 1 µV/cm. Since resistive heating from the contact resistances required fast current ramps with a stepwise current increase of 1 – 2 % of $I_c$, the true $I_c$ value is in general slightly underestimated by choosing the current value one step lower than the first data point in the take-off part of the *U-I* curve as $I_c$-value.

$T_c$ measurements were performed in the same experimental arrangement, applying a resistive 4-point method ($I$ = 5 mA) and measuring the sample resistance during the heat up ramp. Temperature was controlled via a calibrated carbon-glass-resistor in very close neighbourhood to the sample.

# Results

Resistive $T_c$ measurements show that for both kinds of wires nearly optimum values were obtained (see Tab. 2). The only difference is, that the in-situ wires had a somewhat broader transition, probably due to a less homogeneous formation of $MgB_2$. In Fig.3 resistive $T_c$ transitions at zero field and $B = 0.5$ T are given. No broadening of the transitions with field up to 0.7 T was observed. Plotting $T_c$ (midpoint) vs field gives a nearly linear correlation (see Fig.4). We obtained a slope of the graph (evaluated from 0.3 T to 0.7 T) of $dT_c/dB = -3.8$ K/T which extrapolates to an irreversibility field of about 10 T at 4.2 K.

Evaluating the current carrying capacity of the samples, transport critical current measurements were first performed at 4.2 K in the LHe bath. Regarding the results of the wires made from commercial $MgB_2$ powders, we measured 65 kAcm$^{-2}$ at $T = 4.2$ K in self field for both sheath composites with Ta and Nb. For Nb, the self field values are influenced by the current carrying capacity of the Nb layer. But the agreement with the values for the conductor embedded in the Ta sheathing at higher fields and the extrapolation to zero field confirm the results.

In Fig.5, the critical current in dependence on field is given, showing a hysteresis for increasing and decreasing field, probably related to some remanent magnetization. This hysteresis is given in more detail in the insert of fig.5, where a smaller loop was measured for moderate high fields. The reason for this hysteresis is under investigation performing analysis of the microstructure and corresponding magneto-optical imaging (MOI) of the current flow. The maximum of the critical transport current in zero external field, 160 A, corresponds to $J_c = 65$ kAcm$^{-2}$. The corresponding volume pinning force behavior shown in Fig.6 has its maximum at approximately $B = 3.5$ T. The Kramer plot using these data, given in Fig.7, indicates that the irreversibility field $m_0H^*$ (4.2 K) is about 10 T.

For the in-situ wires, the field dependency of the transport critical currents looks quite different. As shown in Fig.8 in comparison to the wires from reacted precursors, the zero field values are significantly higher reaching about $J_c = 100$ kAcm$^{-2}$ at 4.2 K. At higher fields the currents decrease more rapidly. This behaviour is also expressed in Fig.9, the comparison of the volume pinning force. The in-situ wires have their maximum pinning at significant lower fields of about 1 - 1.5 T.

In contrast to the wires from prereacted powders, for in-situ wires the sheath composition obviously plays a role. As shown in Fig.10, a Nb sheath leads to higher $J_c$ values at zero field compared to the wire with a Ta containing sheath. This behaviour is also expressed in the pinning behaviour shown in Fig.11. For Nb containing sheaths, a pinning maximum is found at $B = 1$T, whereas for a Ta containing sheath the maximum shifts to 1.5 T. The explanation of this effect is still unclear. It may result from a different pre-stress state, due to a significant difference in the thermal expansion of Ta and Nb. This has to be proven by stress experiments. However, a different chemistry with a resulting different phase composition and microstructure cannot be excluded as a second reason.

Regarding the temperature dependence of the critical currents of both types of wires as given in Fig.12, we found no differences within the errors. The critical current decreases almost linear with temperature, reaching very small values above 35 K, close to $T_c$. We also investigated the field dependence at temperatures of 20, 25 and 30 K, which is given in Fig.13. In this temperature range, the irreversibility field is quite low, at $T = 20$ K it is close to 3 T. This behaviour was found to be very similar for all wires.

**Discussions and conclusions**

We prepared and investigated two different kinds of wires, each made with two different sheath composites using Nb and Ta as first sheath wall and a mechanical reinforcement with stainless steel. Main results are summarized in Tab. 1. We could achieve very high transport currents in zero external field up to 100 kAcm$^{-2}$ with excellent reproducibility. For the different wire types, the maximum of the volume pinning force is found between 1 T and 3 T. The zero field transport current densities decrease significantly down to about 2/3 of that measured for the Mg/B type wire with a Nb sheathing (65 kA/cm$^{-2}$ compared to 100 kA/cm$^{-2}$). This is an indication, that the pinning behavior was different in the wires as a consequence of the preparation method. Since all wires have been treated together in the same heat treatment program, the precursor and/or the material of the first wall, Nb or Ta, has to be responsible for this effect. So far we have not analyzed the true reason for this observation. As it seems to be the most probable reason, the microstructure of the precursor has to be investigated which is already under work. The following arguments have to be considered. The pre-reacted MgB$_2$ was fine grained with homogeneously distributed MgO impurity. We expect that even after the heat treatment, MgO is still very well distributed and leads to a restriction of grain growth during phase recovery. For the Mg/B precursor, the situation is quite different. MgO is introduced into the filament via the surface oxide layer on Mg particles. During the heating, Mg melts above 650°C and the MgO coexists as secondary phase. In principle, there is no mechanism which distributes MgO homogeneously in the filament. - This hypothesis has to be proved with upcoming microscopic investigations, it could explain perhaps the observed behavior. Regarding also the fact that the heat treatment was not optimised so far, we don't want to speculate too much prior to extended microscopic work about phase quality, purity and grain structure. Especially the presence of a reaction layer at the filament-sheath interface, being more or less pronounced for the two wire types, may additional influence the final phase constitution.

The second possible explanation, the influence of pre-strain effects will be investigated by stress-strain experiments in the near future, an optimized steel content should come out from those data. It is expected that strain effects influence significantly the superconducting properties in wire composites. Here not only induced thermal prestrain has to be considered, in addition also residual stress due to the strong anisotropy in the thermal expansion of the lattice axis a and c, being a factor of 2 [16], are expected in fully dense filaments. This situation is again analogous to that in Chevrel phase wires [19]. - Summarizing the observations, we found that the preparation method can influence strongly the wire properties in zero and finite background field. There is a good chance to find methods for enhancing the pinning in this system.

For possible applications of these wires, we found good indications for further improvements. As a first goal, the irreversibility field has to be improved significantly. For LHe applications the present wires can only be applied up to about 5 – 6 T, and at higher temperatures above 20 K only about 1 T or even smaller fields can be tolerated. A future improvement of the irreversibility field close to $\mu_0 H_{c2}$, however, could bring this new superconductor system much closer to applications in transformers, coils, motors or current limiters in conjunction with the use of cryocoolers.

## Acknowledgement

We would like to thank Dr.W Schauer for critical discussions and valuable suggestions during preparation of the manuscript.

| Material | Linear thermal expansion coefficient at T = 300 K x $10^{-6}$ K$^{-1}$ |
|---|---|
| MgB$_2$ | 8.3* |
| Cu | 16.7 |
| Stainless steel | 18 |
| Fe | 11.8 |
| Ni | 12.8 |
| Nb | 7.1 |
| Ta | 6.5 |

Table 1 : Linear thermal expansion coefficients for selected conductor materials. * was calculated from data in ref. [16]

| Sample number | 1 | 2 | 3 | 4 |
|---|---|---|---|---|
| Precursor | MgB$_2$ | MgB$_2$ | Mg / B | Mg / B |
| Wire composition in %: | | | | |
|   Filament | 20 | 20 | 14.5 | 14.5 |
|   Ta | 13 | | 14.5 | |
|   Nb | | 13 | | 14.5 |
|   Cu | 15 | 15 | 20 | 20 |
|   Stainless steel | 52 | 52 | 51 | 51 |
| Critical temperature $T_c$(50%)[K] | 38.4 | 38.4 | 37.3 | 37.3 |
| Width of $T_c$ transition [K] | 0.5 | 0.5 | 3 | 3 |
| Zero field critical current density $J_c$ [kAcm$^{-2}$] at temperature: | | | | |
| $T$ = 4.2 K | | 65 | 88 | 100 |
| $T$ = 20 K | | 32 | | 52 |
| $T$ = 25 K | | 23.5 | | 37 |
| $T$ = 30 K | | 13.5 | | 23 |
| $F_p^{max}$ (4.2K) at $B$ [T] | | 3.5 | 1.5 | 1 |
| $\mu_0 H^*$ [T] at $T$ = 4.2 K | 10 | 10 | 6-7 | 6-7 |

Table 2

Figure captions:

Fig.1 Cross section of MgB$_2$ monofilamentary wire with 1.18 mm diameter. The composition of the sheath structure is given by: stainless steel / Cu / Ta or Nb, from the outside to the filament. Material contents are given in Tab.1.

Fig.2 Long length sample (1m length, 33 mm diam.) prepared by the wind & react Technique.

Fig.3 Resistive $T_c$ transition in zero field and $B$ = 0.5 T. For detailed data see Tab.1.

Fig.4 Resistive $T_c$ vs. $B$ for a MgB$_2$ wire (midpoint values).

Fig.5 Transport critical current density of MgB$_2$ wire vs increasing and decreasing magnetic background field B. For increasing field, the measurement starts at B = 1 T .

Fig.6 Volume pinning force with field for a MgB$_2$ wire with increasing and decreasing field.

Fig.7 Kramer plot for a MgB$_2$ wire, leading to an irreversibility field of 10 T at 4.2 K.

Fig.8 Critical current density of in-situ MgB$_2$ wires vs field at 4.2 K compared to MgB$_2$ wires.

Fig.9 Volume pinning force of the $I_c$ vs $B$ graphs of Fig.9.

Fig.10 Critical current densities vs field at 4.2 K for in-situ MgB$_2$ wires with Ta/Cu/SS and Nb/Cu/SS sheath.

Fig.11 Pinnig force vs field for the wires and $I_c$ vs $B$ curves of Fig.11.

Fig.12 Temperature dependence of the critical current densities of MgB$_2$ and in-situ MgB$_2$ wires.

Fig.13 Critical currents vs background field at $T$ = 20, 25, 30 K for MgB$_2$ wires.

**References**

[1]   Akimitsu J 2001 Symp. on Transition Metal Oxides, 10 January 2001 (Sendai)
[2]   Nagamatsu J, Nakagawa N, Muranaka T, Zenitani Y and Akimitsu J 2001 *Nature* **410** 63
[3]   Bud'ko S L, Lapertot G, Petrovic C, Cunningham C E, Anderson N and Canfield P C 2001 *Phys. Rev. Lett.* **86** 1877
[4]   Larbalestier D C, Cooley L D, Rikel M, Polyanskii A A, Jiang J, Patnaik S, Cai X Y, Feldmann D M, Gurevich A, Squitieri A A, Naus M T, Eom C B, Hellstrom E E, Cava R J, Regan K A, Rogado N, Hayward A, He T, Slusky J S, Khalifah P, Inumaru I and Haas M 2001 *Nature* **410** 186
[5]   Finnemore D K, Ostenson J E, Bud'ko S L, Lapertot G and Canfield P C 2001 *Phys. Rev. Lett.* **86** 2420
[6]   Canfield P C, Finnemore D K, Bud'ko S L, Ostenson J E, Lapertot G, Cunningham C E and Petrovic C 2001 *Phys. Rev. Lett.* **86** 2423
[7]   Patnaik S, Cooley L D, Gurevich A, Polyanskii A A, Jiang J, Cai X Y, Squitieri A A, Naus M T, Lee M K, Choi J H, Belenky L, Bu S D, Letteri J, Song X, Schlom D G, Babcock S E, Eom C B, Hellstrom E E and Larbalestier D C 2001 *Supercond. Sci. Technol.* **14** 315
[8]   Sumption M D, Peng X, Lee E, Tomsic M and Colling E W 2001 Preprint cond-mat/0102441
[9]   Canfield P C, Finnemore D K, Bud'ko S L, Ostenson J E, Lapertot G, Cunningham C E and Petrovic C 2001 *Phys. Rev. Lett.* **86** 2423
[10]  Jin S, Mavoori H and van Dover R B, submitted to *Nature*
[11]  Glowacki B A, Majoros M, Vickers M, Evetts J E, Shi Y and McDougall I 2001 *Supercond. Sci. Technol.* **14** 193
[12]  Grasso G, Malagoli A, Ferdeghini C, Roncallo S, Braccini V, Cimberle M R, and Siri A S, submitted to *Superc. Sci. Technol.*
[13]  Kang W N, Kim H-J, Choi E-M, Jung C U and Lee S-I 2001 Preprint cond-mat/0103179
[14]  Eom C B, Lee M K, Choi J H, Belenky L, Song X, Cooley L D, Naus M T, Patnaik S, Jiang J, Rikel M, Polyanskii A, Gurevich A, Cai X Y, Bu S D, Babcock S E, Hellstrom E E, Larbalestier D C, Rogado N, Regan K A, Hayward M A, He T, Slusky J S, Inumaru K, Haas M K and Cava R J 2001 *Nature* at press
[15]  Fluekiger R, Goldacker W, *IEEE Trans. Magn. MAG-21*, 1985, **907**
[16]  Jorgensen J D, Hinks D G, Short S, *Physical Review* B, 2001, 63 224522
[17]  Rimikis G, Goldacker W, Specking W and Flükiger R 1991 *IEEE Trans. on Magn.* **27** 1116
[18]  Goldacker W, Specking W, Weiss F, Rimikis G and Flükiger R, 1989 *Cryogenics* **29** 955
[19]  Goldacker W, Miraglia S, Hariharan Y, Wolf T, Fluekiger R, *Adv. Cryog. Eng. Mater.,* 1988, **34** 655

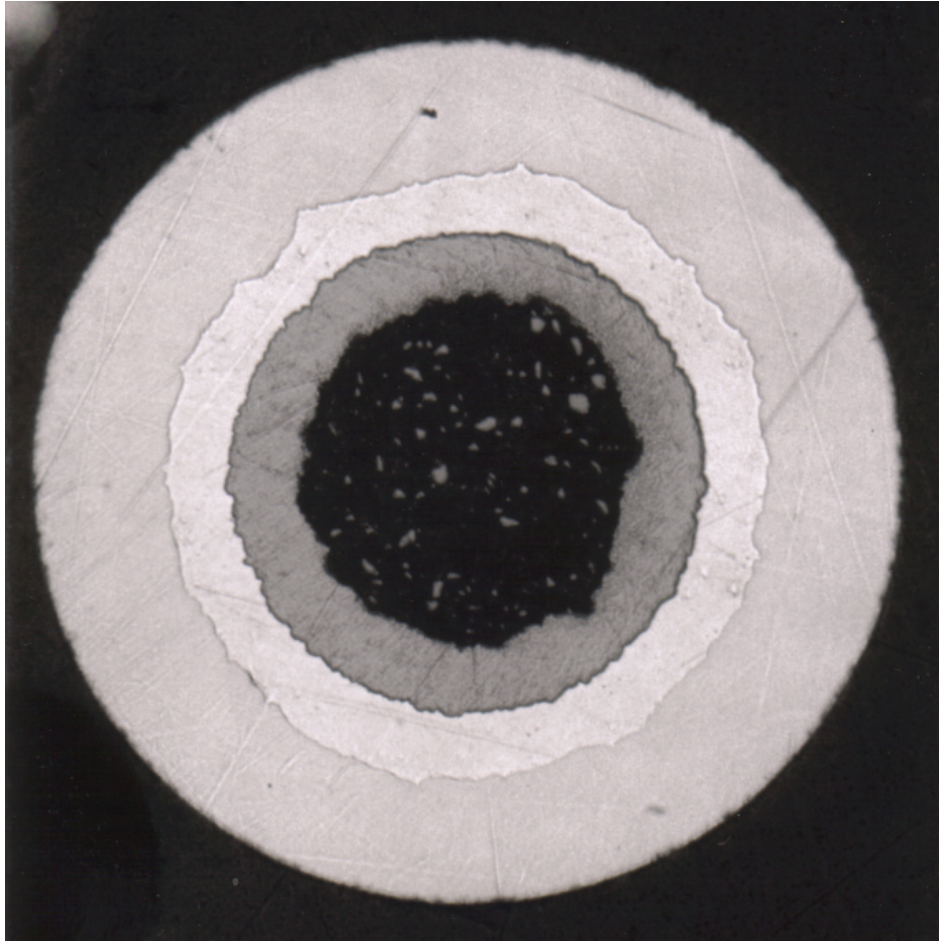

Fig.1

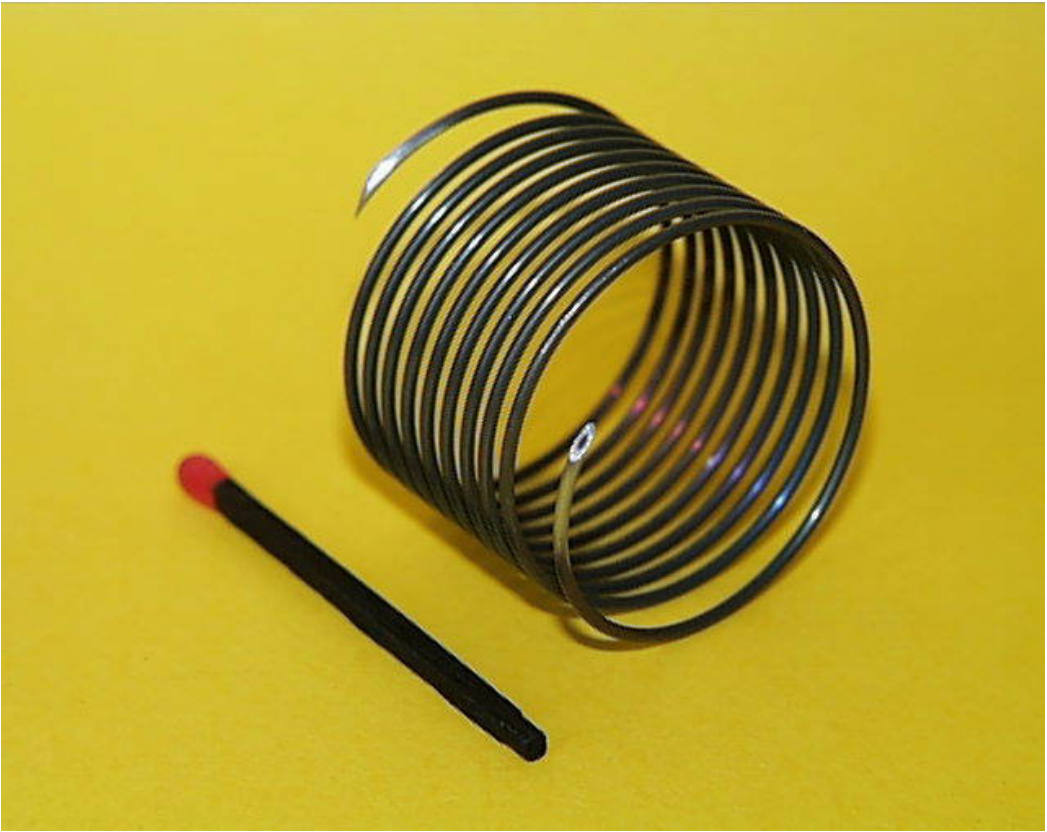

Fig.2

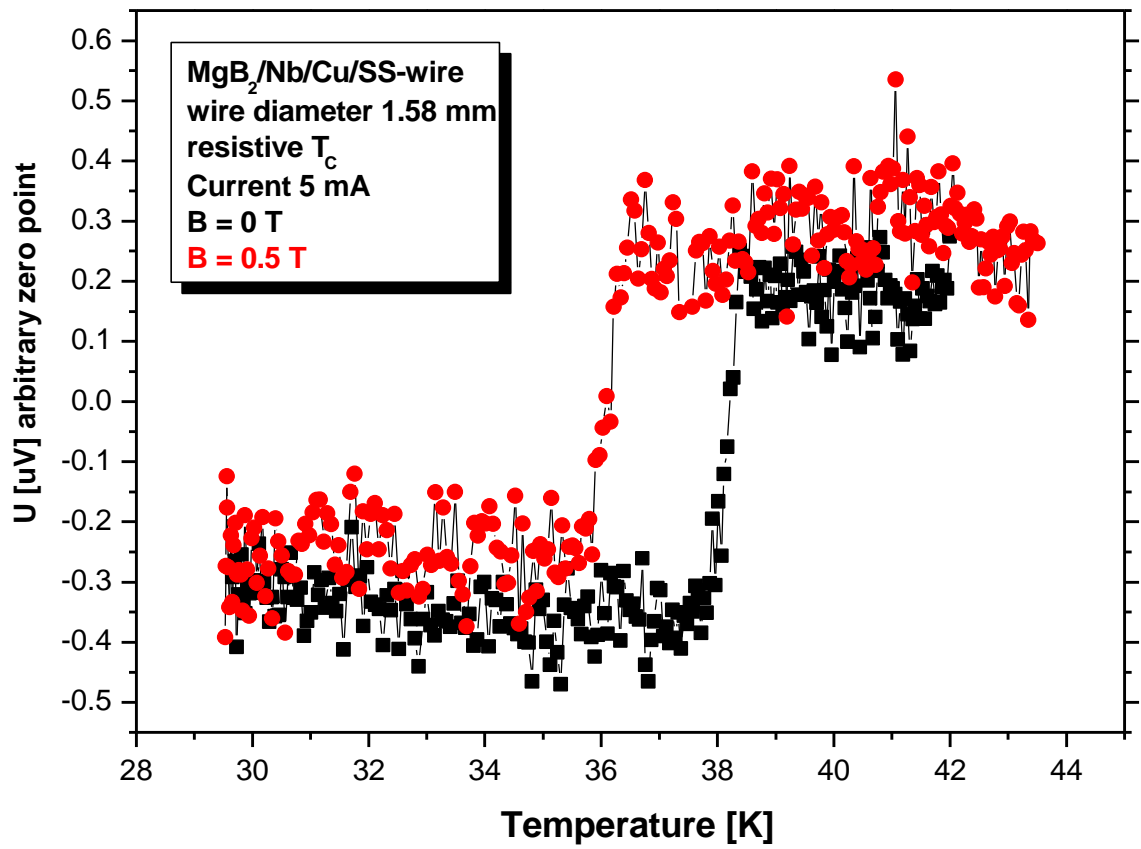

Fig.3

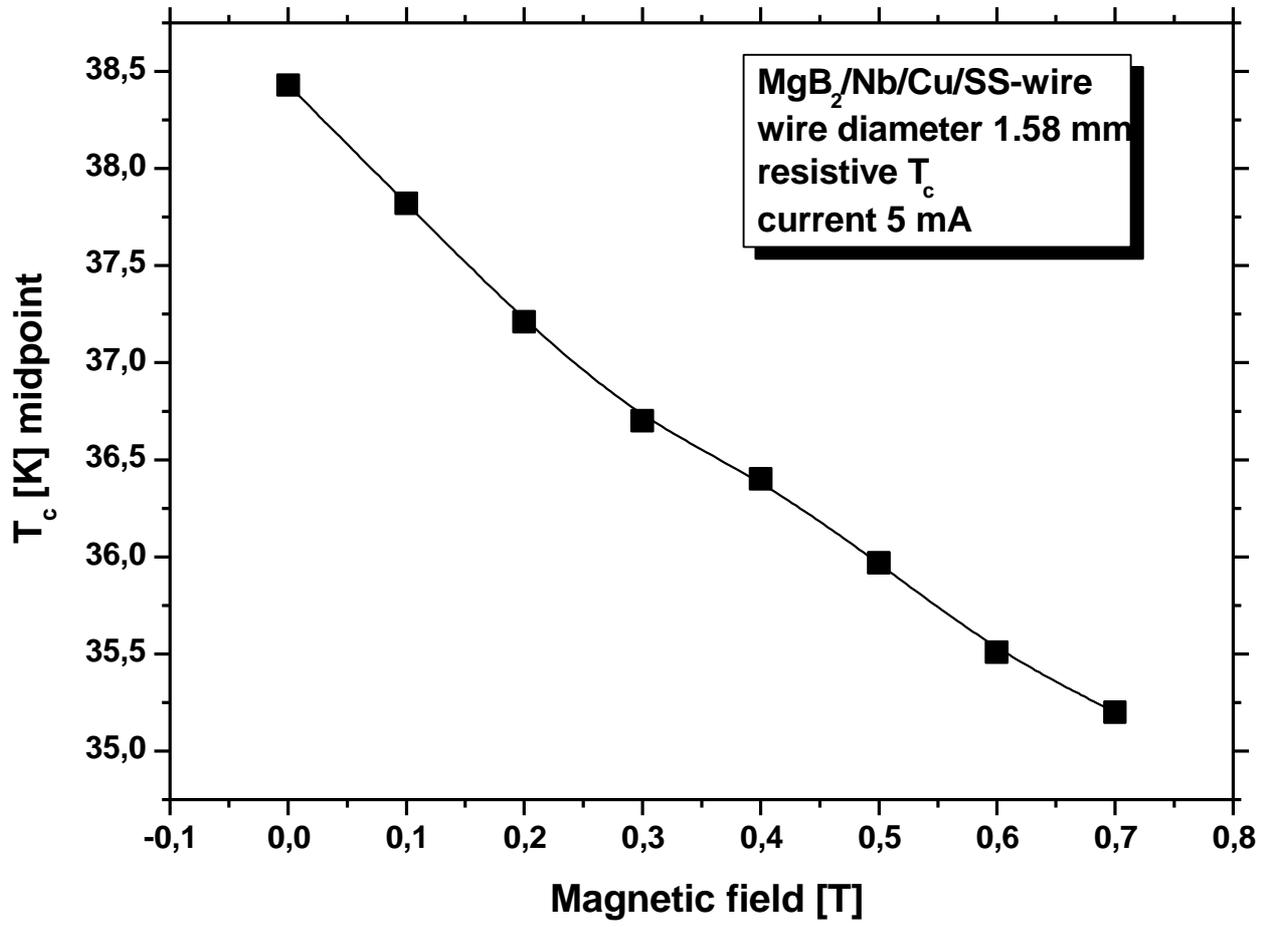

Fig.4

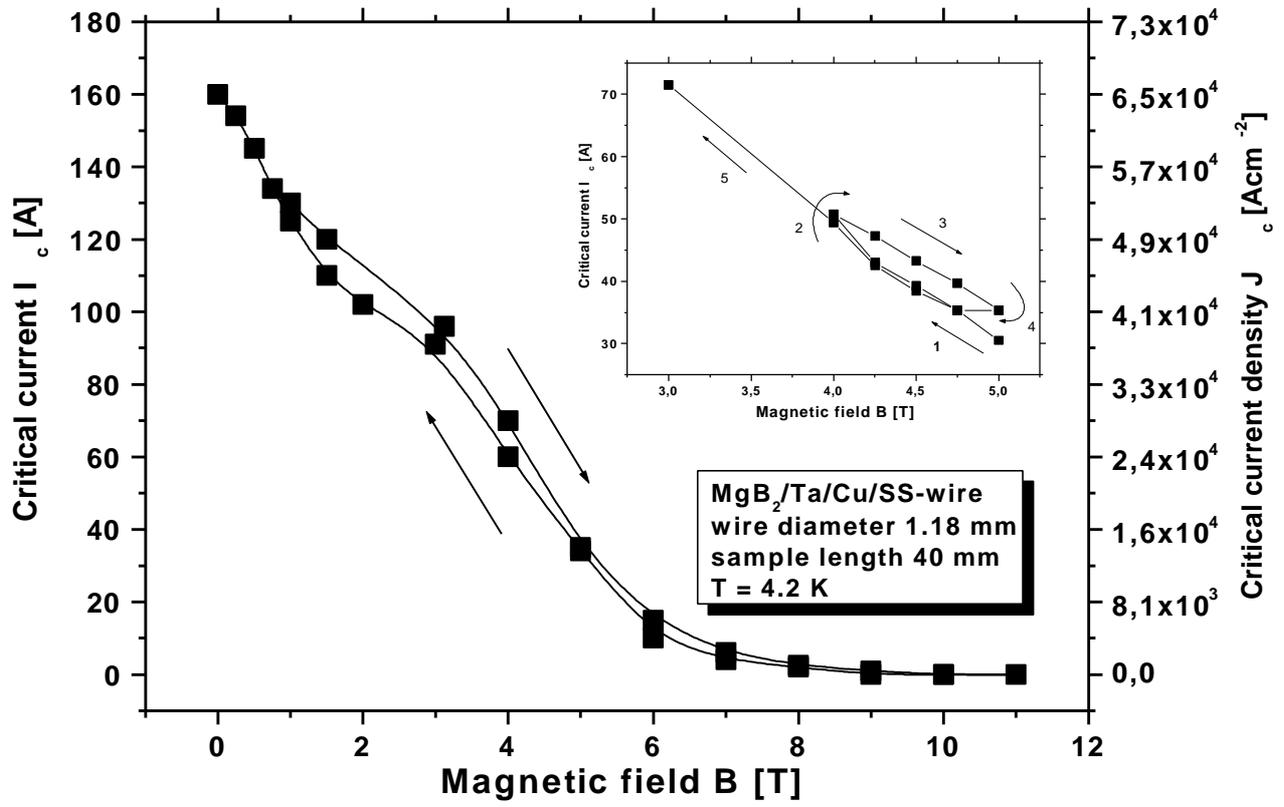

Fig.5

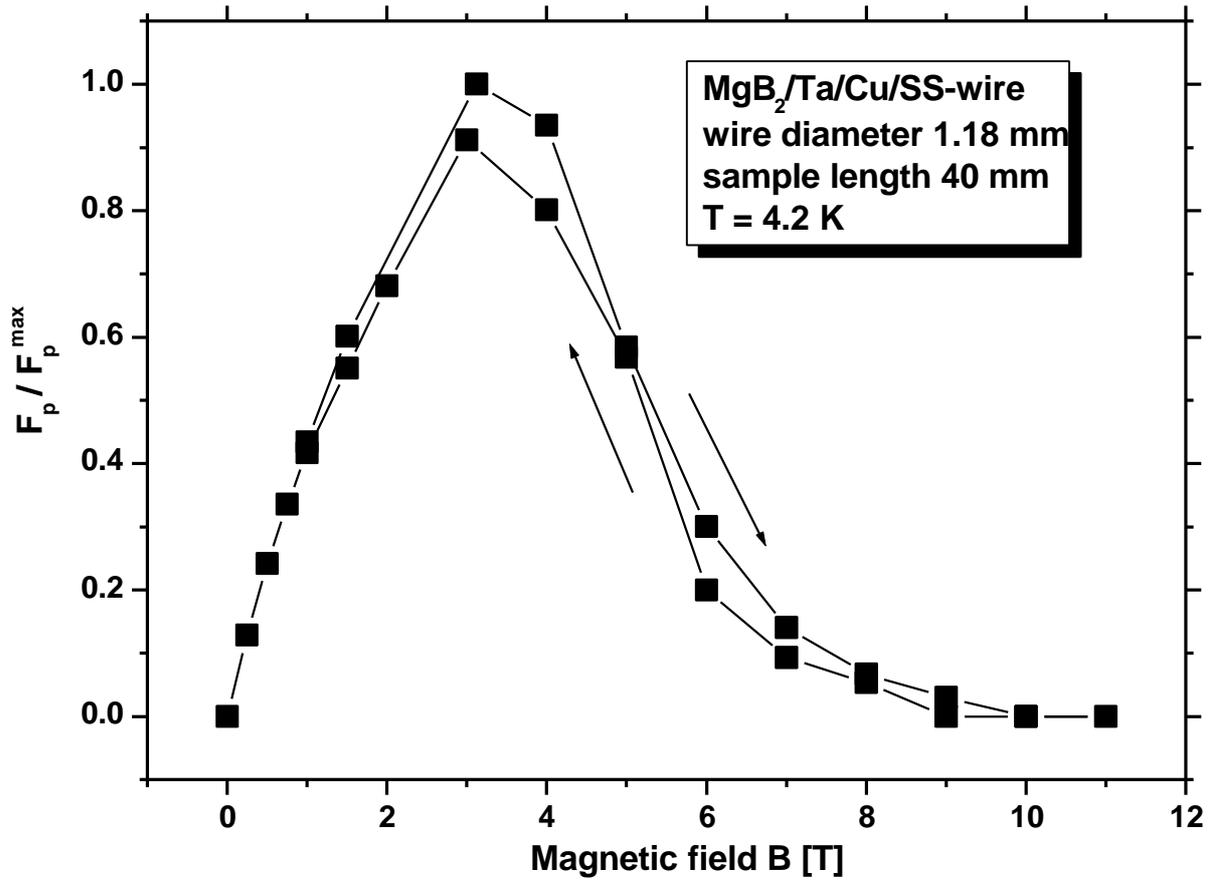

Fig.6

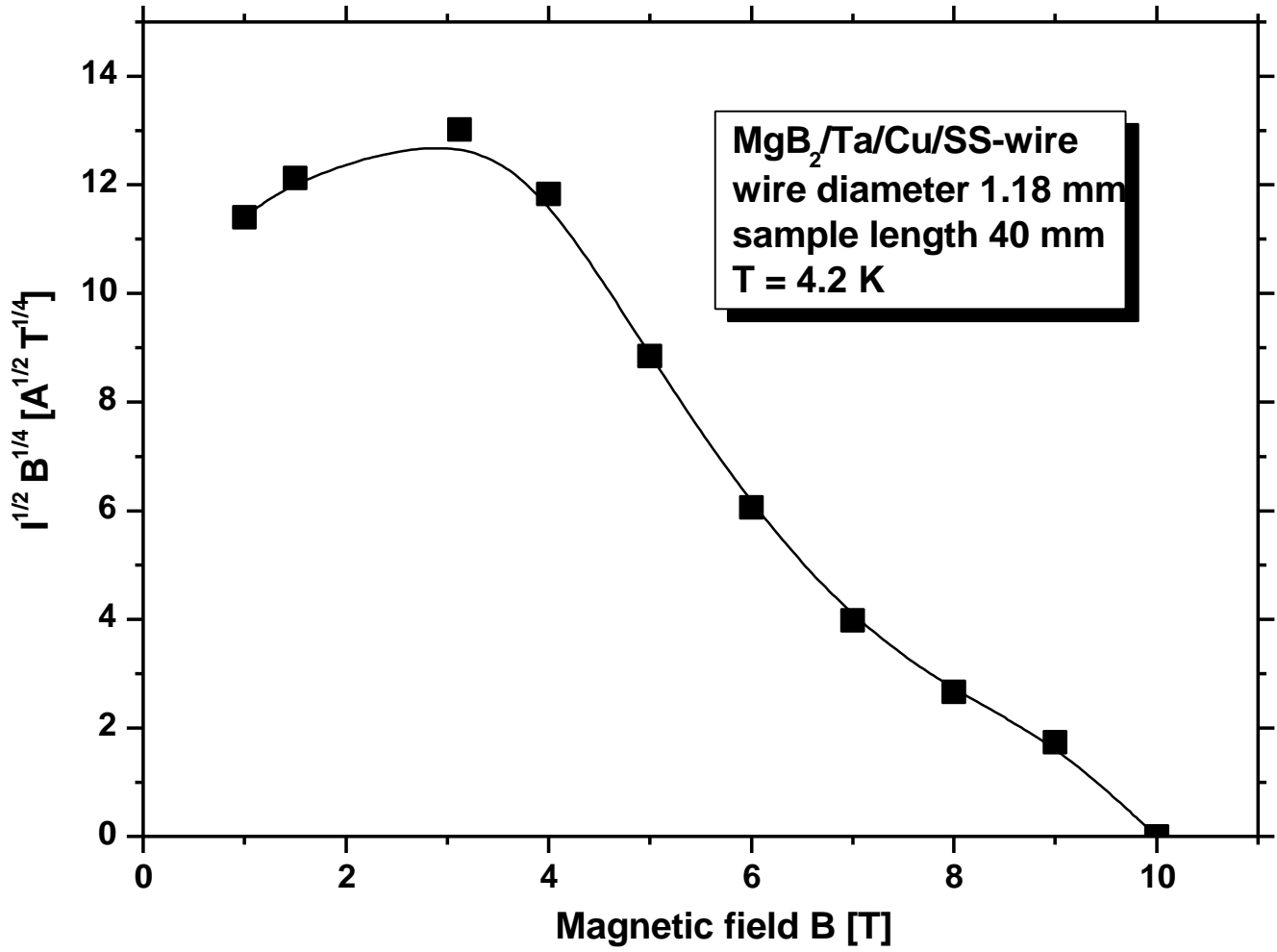

Fig.7

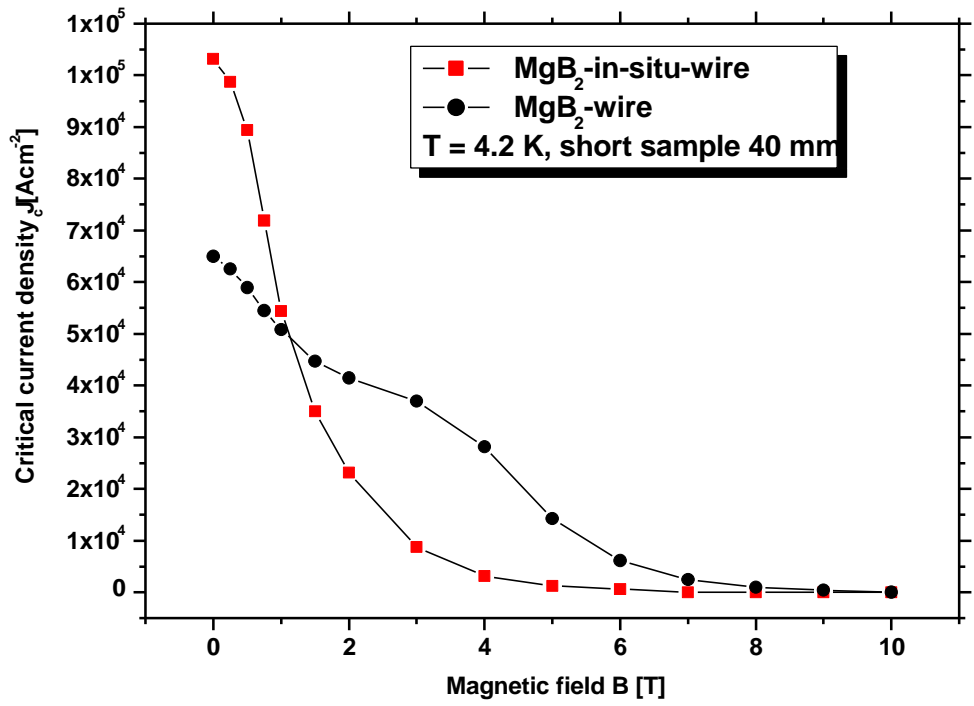

Fig.8

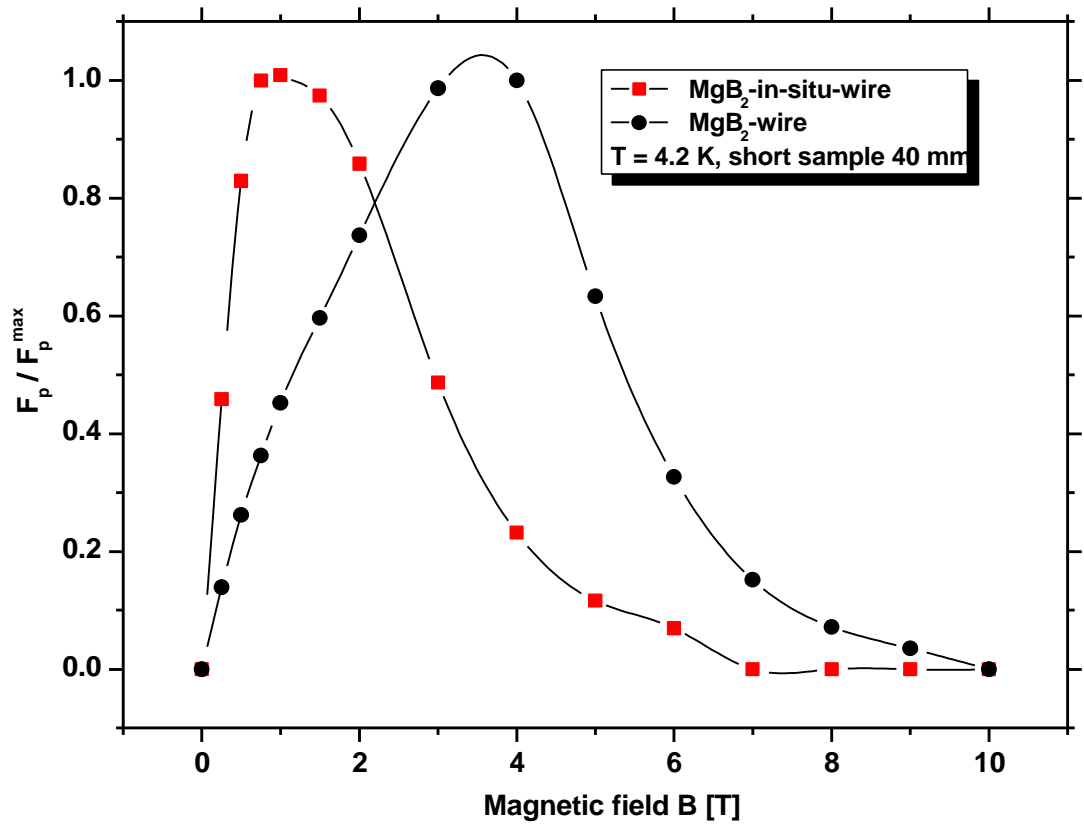

Fig.9

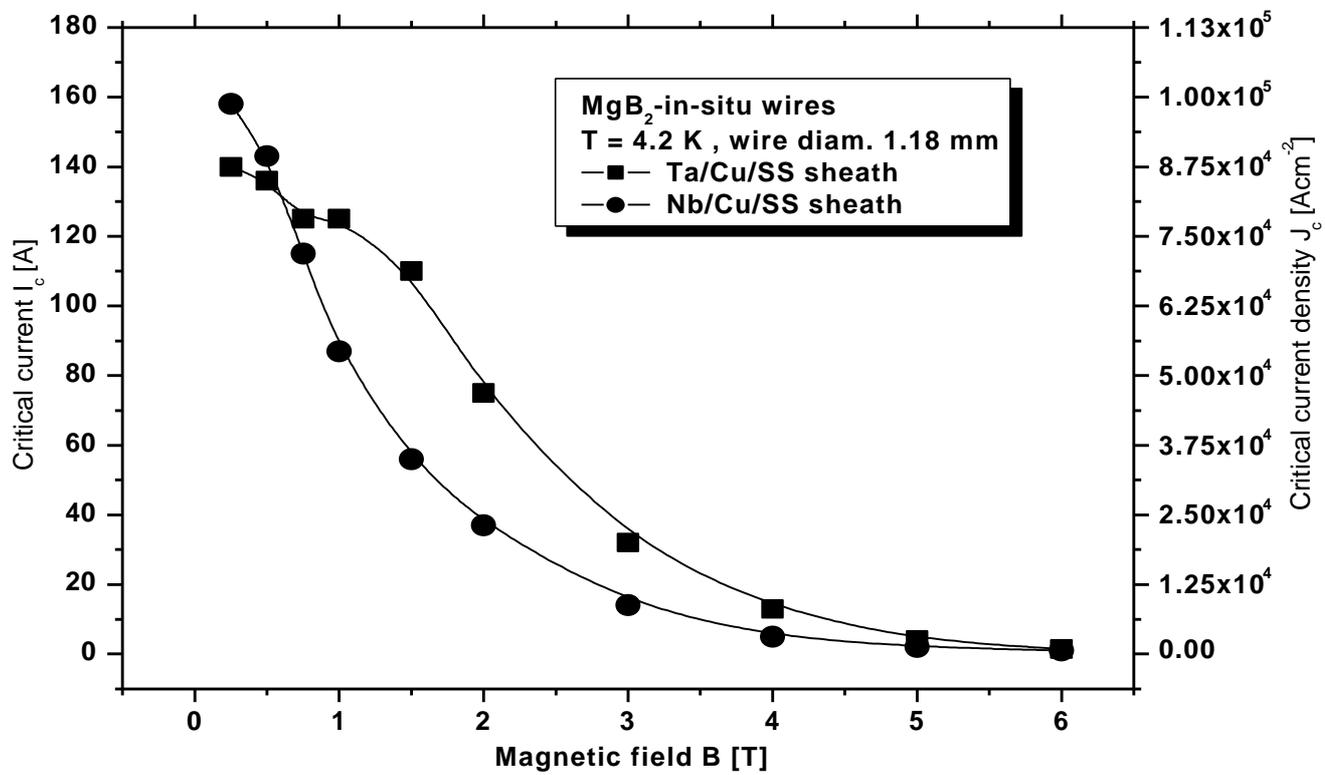

Fig.10

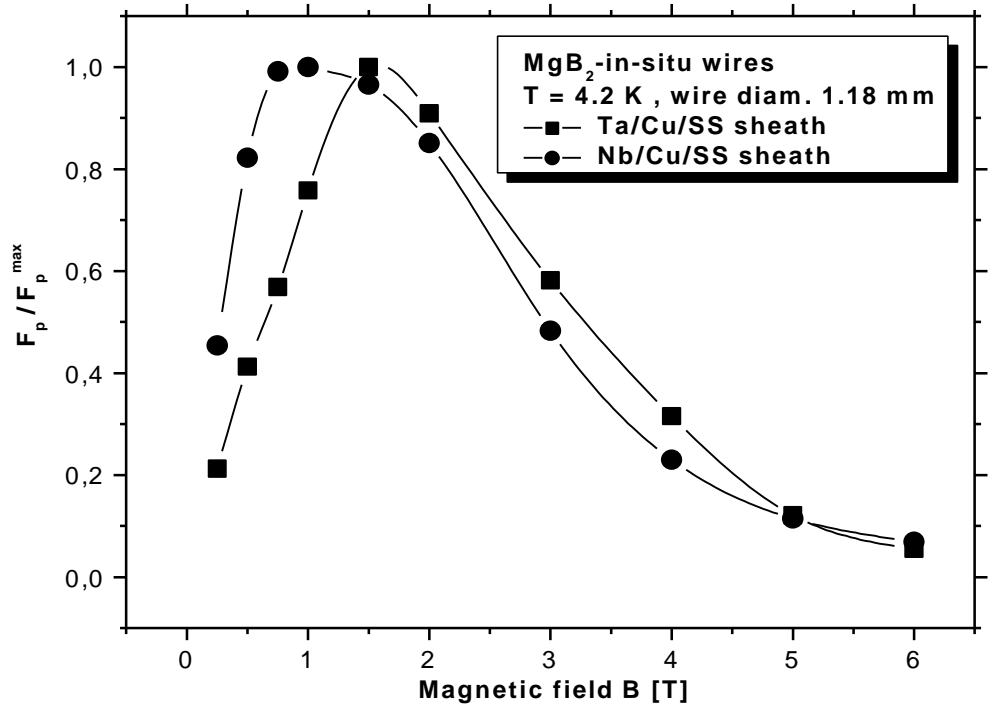

Fig.11

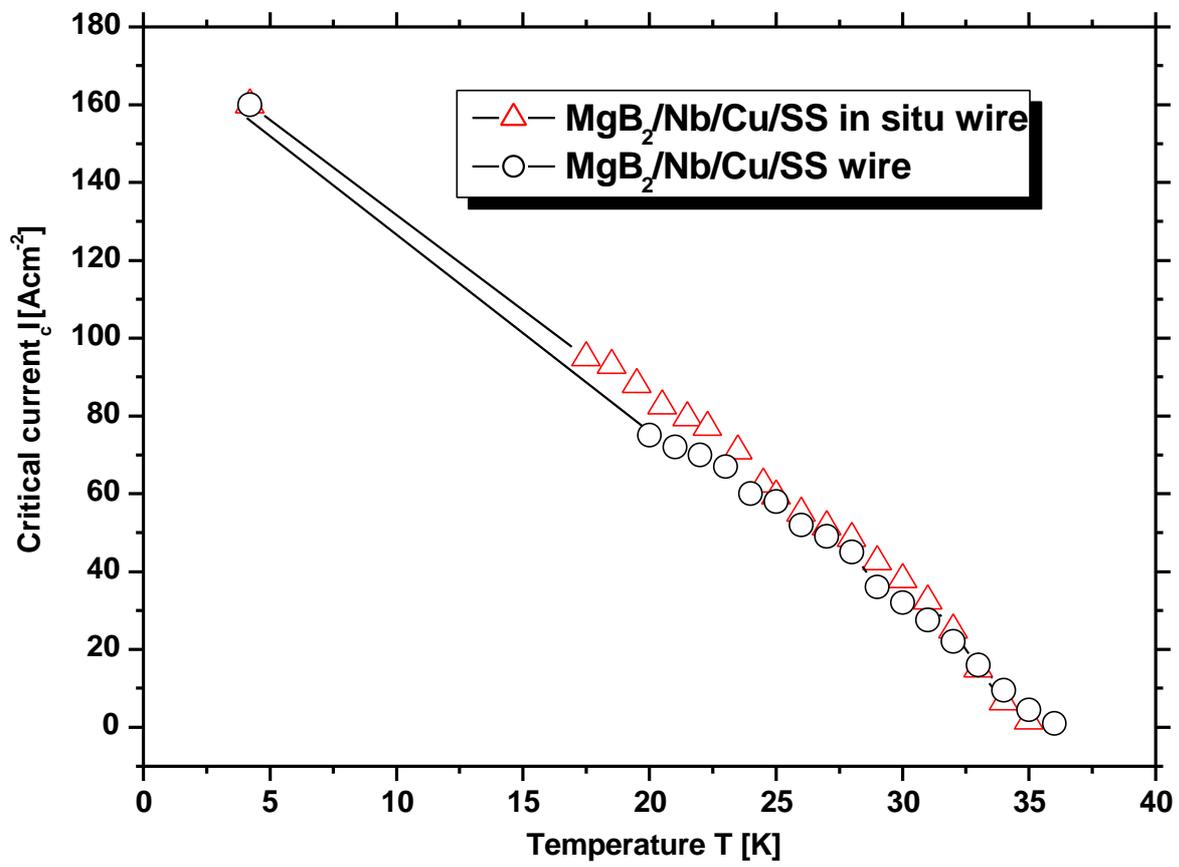

Fig.12

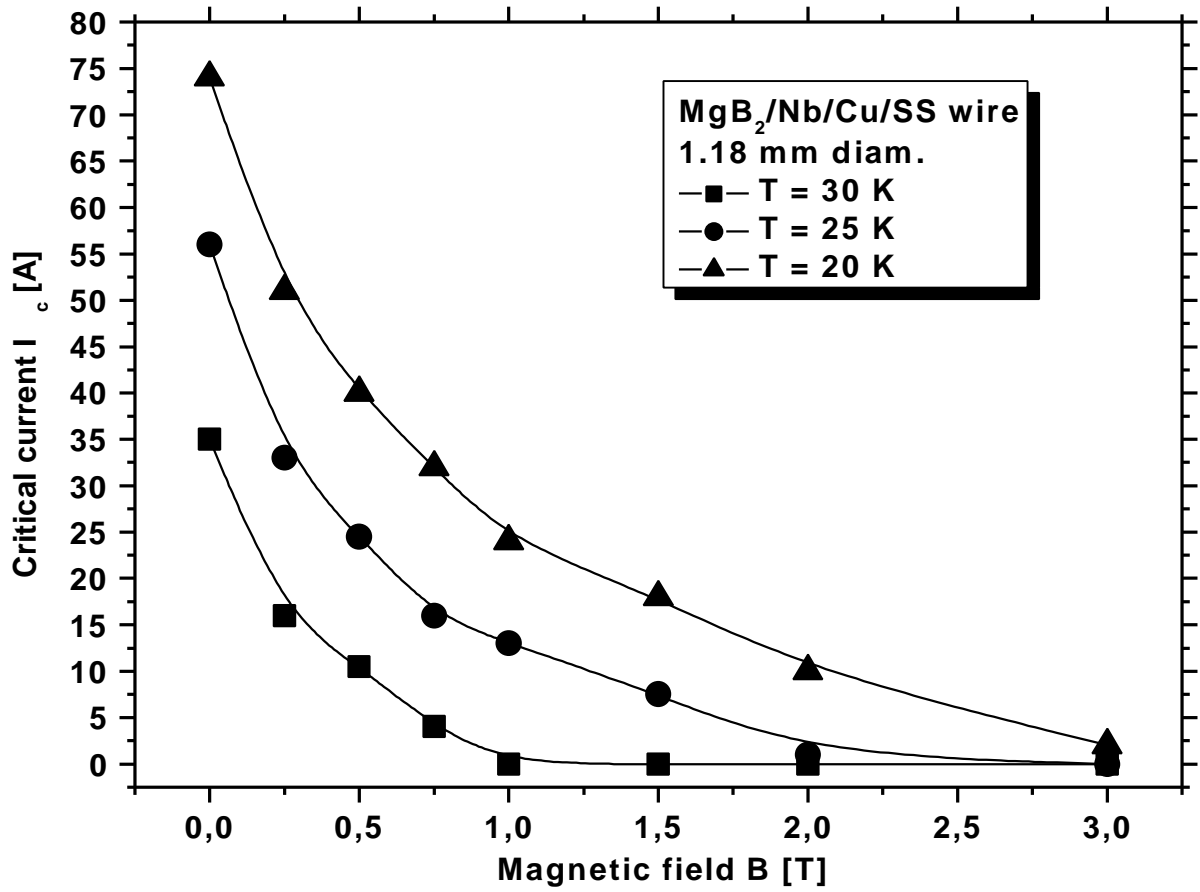

Fig.13